# Egg yolk as a model for gelation: from rheometry to flow physics

Maxwell C. Marsh[1], Mohammad Tanver Hossain[1], Randy H. Ewoldt[1]

[1]Department of Mechanical Science and Engineering, University of Illinois, Urbana-Champaign

## Abstract

Egg yolks are an excellent model for studying sol-gel transitions, particularly the power law viscoelasticity that defines the critical point of gelation. However, prior studies lack comprehensive datasets and fail to visualize flow behavior linked to temperature and time-dependent linear and nonlinear rheology. Here we present a detailed dataset characterizing egg yolk viscoelasticity across temperature, time, and forcing amplitude using oscillatory shear, step strain, step stress, and constant high strain rate. Novel protorheology visualizations link rheological properties with observable flow behavior. Our findings highlight the nuanced determination of the critical gel point, emphasizing observation timescale dependencies. We compare methods to identify critical temperatures for gelation, including power law viscoelasticity, moduli crossover, diverging zero-shear viscosity, and emerging equilibrium elastic modulus, while visualizing flow consequences near these transitions. Egg yolk is an accessible non-toxic material relevant to the physicist and the chef alike, making it ideal for understanding the rheology of critical gels. By integrating protorheology photos and videos with rigorous rheometric data, we deepen the understanding of critical gels, with broader impacts for teaching and modeling sol-gel transitions.

# I. Introduction

Gelation is a critically important phenomenon to materials processing in manufacturing[1,2], food[3,4], health science[5,6], and many other domains[7]. The transition from a liquid solution to a solid gel (the sol-gel transition) may involve the identification of a specific "gel" temperature, well known to be characterized by a diverging zero-shear viscosity and an emerging equilibrium elastic modulus. This asymptotic behavior is also consistent with power-law viscoelastic relaxation in the terminal viscoelastic regime[8–11]. Rheological properties near this critical gel condition are interesting, in terms of linear and nonlinear rheometric material functions, as well as the consequences on macroscopic non-trivial flow scenarios.

Among the many materials used to understand the physics of the sol-gel transition, egg yolks are a particularly desirable model material. They are universally accessible, non-toxic, and familiar, making them ideal for understanding the rheology of critical gels, including power law viscoelasticity that defines the critical point of gelation. While excellent prior studies have reported power law viscoelasticity and critical gel temperatures for native [12–14] and modified egg yolk[15–19], they lack comprehensive rheometric datasets with a range of material functions across a range of timescales and forcing amplitudes, and they do not contain visualization (photos and videos) of the consequent flow behavior linked to temperature- and time-dependent linear and nonlinear rheology. Alternatively, excellent visualizations of the temperature-dependence of egg yolks are prevalent in the culinary world but these typically lack the detailed rheological understanding of gelation dynamics and flow consequences[20–24].

Here, we present a detailed dataset characterizing egg yolk viscoelasticity across temperature, time, and forcing amplitude using oscillatory shear, step strain, step stress, and constant strain rate. Novel protorheology visualizations link rheological properties with observable flow behavior. Our findings highlight the nuanced determination of the critical gel point, emphasizing observation timescale dependencies. We compare methods to identify critical temperatures for gelation, including power law viscoelasticity, moduli crossover, diverging zero-shear viscosity, and emerging equilibrium elastic modulus, while visualizing flow consequences near these transitions. By integrating protorheology photos and videos with rigorous rheometric data, we deepen the understanding of critical gels, with broader impacts for teaching and modeling sol-gel transitions.

# II. Background

Gelation of protein-laden liquids is commonly classified into two categories, i.e. physical and chemical[25], and sometimes a third, biochemical[26]. For egg yolks, heat-induced gelation is a physical process and the most widely encountered. Egg yolk is composed of granules inside a continuous plasma phase which corresponds to about 78% of yolk dry matter. The plasma, which is liquid at room temperature, contains roughly 85% low-density lipoproteins (LDL) and 15% livetin proteins. The remaining 22% of dry yolk matter are granules which are primarily made up of high-density lipoproteins (HDL)[27]. It is generally accepted that upon application of heat, the LDLs will undergo what is called "cluster fusion" and form large aggregates which eventually lead to a large 3D network[14].

Gelation is evidenced by physical (rheological) properties: a diverging zero-shear viscosity and the emergence of an equilibrium elastic modulus, as in the Figure 1 schematic for heat-induced gelation adapted from Winter and Chambon (1986)[8]. Photos in Figure 1 show familiar observable flow consequences across this gelation transition for egg yolks at three different temperatures: heated yolks were held in a temperature-controlled water bath for 60 min then placed on a flat surface and cut, a type of *protorheology*[28] experiment to intuitively reveal relevant rheological properties. At 25 °C, the yolk flows like a liquid after the membrane is cut (photo taken 10 seconds after cut). At 75 °C, the yolk holds its shape like a solid and exhibits brittle-like fracture[29]. At an intermediate temperature of 67 °C, near the "sexy egg" condition known by the culinary world[30], the yolk appears somewhere in between solid and liquid[23,24]: it appears soft and ductile, holding an overall shape like a solid in resistance to gravitational and surface tension forces, but with a smooth surface suggestive of localized liquid behavior (photo taken 10 seconds after cutting the yolk). This heat-induced transition occurs due to microstructural network formation, but it is the asymptotic rheological properties that define the *critical gel point* indicating the transition from a liquid to a solid[8] at these observation timescales.

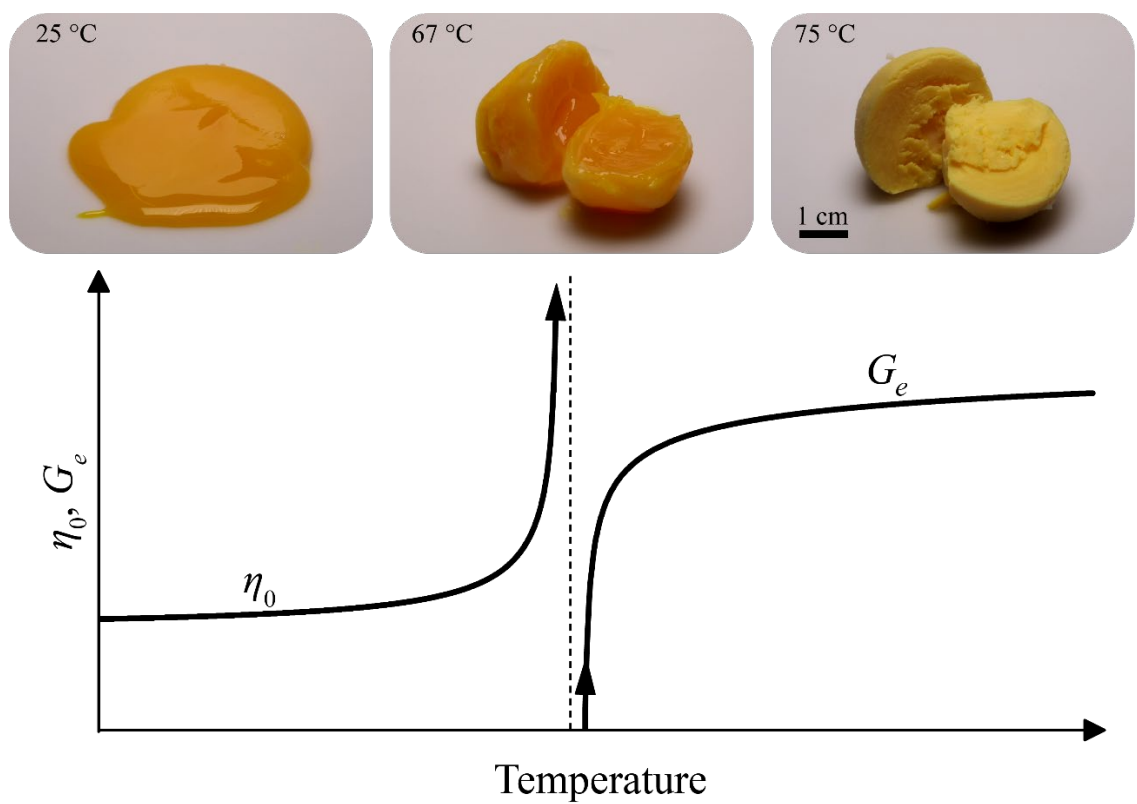

**Figure 1.** Evidence of gelation with increasing temperature: protorheology images and schematic of underlying rheological properties: namely diverging zero-shear viscosity and emerging equilibrium modulus. Pictures of egg yolk at 25 °C, 67 °C, and 75 °C show the transition through the critical gel point. The heated yolks were held in the water bath set to 67 °C and to 75 °C respectively for 60 min each. All pictures were taken ~10 seconds after the yolk was cut.

In practice, the determination of the gel point is obscured by our inability to observe infinite times. Any reported critical temperature for solidification must also report the observation timescale: it is a fundamental tenet of rheology and viscoelasticity that "everything flows" for sufficiently long observation time[31–33]. In addition to viscoelastic time-dependence, egg yolks require a secondary consideration due to background mutation of the microstructure at fixed temperature, which conflates with our finite times of viscoelastic observation. Therefore, multiple potential values of a critical temperature can be defensible as a way to describe the gel temperature with the caveat that it may depend on one's timescale of observation. In this work we compare various methods to determine the gel point and address some of the challenges previously mentioned.

## III. Materials and Methods

### A. Materials

Grade A large white chicken eggs were purchased from a local grocery store, stored in a lab refrigerator, and used within two weeks of purchase. To extract the yolk, we followed the procedure of Harrison and Cunningham (1986)[34]. After separating the yolk and the white by hand, we placed the vitelline membrane on a paper towel to remove any of the thick albumen still left after the separation process. The membrane was then pierced to allow the yolk to collect in a weigh boat while taking care to avoid collecting any of the white or the chalazae. To minimize variation among eggs, three yolks were mixed before being tested.

### B. Temperature controlled water bath

For visualization of whole yolk and whole eggs, raw eggs were heated to precise temperatures (1 °C increments between 55 °C and 75 °C) using a Thermo Fisher Scientific Versacool temperature-controlled water bath. This is analogous to a sous vide heater which is commonly used to cook food to precise temperatures. When only the yolk was desired, the white could be removed and the yolk carefully placed in the bath as shown in Figure 1. However, while conducting protorheology tests seen in Figure 2, it was more practical to place yolk into a 20 mL glass vial and that vial into the water bath. When collecting whole images of eggs as seen in Figure 3, the naturally protective shells meant the entire egg could be placed into the water bath.

### C. Rheometry

Rheological measurements in controlled simple shear deformation were performed on a rotational stress-controlled rheometer (DHR-3, TA Instruments, USA) and a rotational strain-controlled rheometer (ARES-G2, TA Instruments, USA) both equipped with Peltier heating systems. Concentric cylinders (28 mm diameter bob and 30 mm cup) were used on both instruments. Double walled concentric cylinders (28 mm inside cup diameter, 29 mm outside bob diameter, 30 mm inside bob diameter, and 34 mm outside cup diameter) were used only for stress-relaxation experiments. To minimize sample evaporation or coagulation, mineral oil was floated on the surface of the egg yolk[11]. Prior studies on egg yolk have used cone-plate and parallel plate geometry[12,17,18,35,36]. However, during our preliminary tests we observed the formation of a "skin" around the circumference at the outer sample interface, even at ambient temperatures and with the addition of a coating of mineral oil. This led to a false torque reading motivating our use of concentric cylinder geometry which places the sample interface above the primary shear zone, thereby minimizing artifacts from the interface[37–39].

Here we document the experimental procedures for each of the tests conducted in this study. To determine the mutation time of the sample, we loaded the egg yolk into the cup at ambient conditions before ramping up the temperature at 5 °C/min to 59 °C and 65 °C separately. After the target temperature was reached, we imposed an oscillation at 10 rad/s and a strain amplitude of 1 % (previously determined to be within the linear regime). To obtain viscoelastic measurements across the range of temperatures, we loaded the egg yolk into the cup at ambient conditions before raising the temperature to 55 °C. Preliminary tests revealed that at 55 °C, the sample would not mutate over an extended time period. After equilibrating at 55 °C for 10 minutes, we conducted a frequency sweep from 100 rad/s to 0.1 rad/s at a strain amplitude of 1 %. This entire frequency sweep lasted approximately 410 s. After the test concluded, we increased the temperature by 1 °C, allowed the sample to equilibrate for 10 minutes, and repeated the frequency sweep following the procedure from Laca et. al (2011)[12,40]. These frequency sweeps allowed us to plot moduli as a function of temperature for each frequency, moduli as a function of frequency for each temperature, and loss tangent as a function of temperature for each frequency. For stress-relaxation, creep-compliance, and flow sweep tests, we used the same heating scheduling described above. To determine the equilibrium modulus from stress-relaxation tests in the double wall concentric cylinder geometry, at each temperature, we applied a step strain of 1 % for 300 s. To determine the equilibrium modulus and zero-shear viscosity from creep-compliance tests, for temperatures below 68 °C, we applied a step stress of 0.1 Pa. For temperatures at and above 68 °C, we applied a step stress of 2 Pa. These values were determined to be within the linear regime and large enough to produce a sufficient strain. To determine the zero-shear viscosity and nonlinear regime from flow sweep tests, we performed a strain rate ramp up from $10^{-2}$ s$^{-1}$ to $10^{2}$ s$^{-1}$ over a duration of 150 seconds immediately followed by a strain rate ramp down from $10^{2}$ s$^{-1}$ to $10^{-2}$ s$^{-1}$ over a duration of 150 s. To measure the nonlinear oscillatory response near the critical gel point we maintained a fresh sample at 65 °C while imposing a small amplitude oscillation. When the dynamic moduli were comparable to those determined to be near the gel point, we conducted a strain amplitude sweep at a fixed frequency of $\omega$ = 10 rad/s from 0.1% strain to 1000% strain and then back down to 0.1% again to assess reversibility.

## IV. Results and Discussion

### A. Protorheology

The term *protorheology* (from Latin: proto- meaning "first") was recently introduced to describe the paradigm of rapidly inferring rheological behavior from simple tests[28,41,42]. This technique is being increasingly used for qualitative corroboration, quantitative inference, high throughput testing[43], and to extend the range of traditional rheometry[44–46]. Food materials are particularly accessible to protorheology investigation. As humans, we seem particularly good at playing with our food. We squish, stretch, and poke what we make and eat to understand its behavior. These actions generate complex flows which, over years of interacting with food, provide an intuitive understanding of the mechanics of soft materials when deformed under such conditions. Here we aim to connect such accessible interactions with rigorous quantitative viscoelastic properties as a function of time, temperature, and forcing strength.

In this work, we report a range of protorheology visualizations which are meaningful because they help interpret and quantify what rheological properties we see when we interact with this material in different flow scenarios. This enables us to construct a deeper understanding of complex rheological behavior, including what it means to be at the critical gel point in terms of observable flow physics. Observations below and above the gel point are also important for context to understand what is distinct at the critical gel point.

Figure 2 shows a diverse set of protorheology tests below and above the gel temperature (25 °C and 75 °C). Figure 2(a) – (c) at 25 °C demonstrate (a) finite shear viscosity from flow down an incline, (b) liquid behavior as the fluid settles to the bottom and takes the shape of the tilted vial, and (c) extensional viscosity from capillarity-driven breakup of a fluid filament. Figure 2(d)-(e) at 75 °C post gelation demonstrate (d) viscoelastic energy storage and loss from a bounce test and (e) solid-like behavior from a high lower-bound estimate of viscosity with immobilized material at the top of a tilted vial. These flows are specifically selected to be accessible to the reader to make their own observations. Quantitative properties can be inferred and associated with the particular observation *timescale* and *forcing strength* conditions (e.g., flows may not be steady state or in the linear viscoelastic regime)[28].

An inferred shear viscosity $\eta < 6$ Pa.s is associated with the flow in Figure 2(a), with images taken 5 seconds apart on an incline at 15°. With an average characteristic thickness of around 0.3 cm, the yolk travels 3.9 cm in 20 s. Assuming a density close to that of water ($\rho = 1000$ kg/m$^3$), neglecting surface tension physics and other interfacial film effects that slow the flow results in an *upper-bound estimate* on viscosity[28] as $\eta = \rho g h^2 \sin\theta / (2 v_s) \approx 6$ Pa.s (with yolk height, $h$, inclination angle, $\theta$, and free surface velocity, $v_s$), associated with a gravitational forcing stress $\sigma \approx \rho g h \sin\theta \approx 8$ Pa and observation time $t = 20$ s. This viscosity is also consistent with the level surface observed in Figure 2(b) easily achieved after 10 min in the tilted vial, although we also observe an uneven meniscus suggestive of contact line pinning and interfacial effects that likely impede the flow of the inclined plane, reinforcing that the viscosity estimate is an upper-bound, $\eta < 6$ Pa.s which is lower than that of honey[47]

An extensional viscosity $\eta_E \approx 40$ Pa.s is implied by the flow in Figure 2(c), from pulling a fork (1.9 cm distance between tips of outermost tines) out of a bath of yolk, forming a liquid bridge. The initially ~2 mm wide bridge thins and eventually breaks after 2.15 seconds. Assuming the viscosity rather than fluid inertia sets the longest time of breakup (large Ohnesorge number $\text{Oh} = \eta_E / \sqrt{\rho \Gamma R} \gg 1$), neglecting gravitational drainage (low Bond number $Bo = \rho g R^2 / \Gamma \ll 1$), and assuming a perfectly cylindrical fluid filament ($X = 1$) consistent with the photos [48], the upper bound estimate of an average extensional viscosity[28] is $\bar{\eta}_E = (2X - 1) \Gamma t_{\text{break}} / D_0 \approx 40$ Pa.s, where the surface tension is a value $\Gamma \approx 40$ mN/m [49]. This is associated with the observation time of 2.15 s, a capillary forcing stress of 40 Pa, and a characteristic extensional strain rate of around 1 s$^{-1}$. As a consistency check, we find Oh = 200 and Bo = 0.25. Thus, gravity plays some small role in the flow. For example, gravitational drainage from the upper reserve liquid can extend the lifetime of thin liquid filaments[50], making the viscosity appear higher than it truly is. Therefore, a reasonable upper-bound estimate of the extensional viscosity is $\eta_E \approx 40$ Pa.s. Comparing extensional and shear viscosities gives an apparent Trouton ratio $\text{Tr} = \eta_E / \eta \approx 7$ at room temperature. Because $\text{Tr} > 3$, this protein-based liquid is extensional thickening which is expected due to stored elastic stresses during extensional flow. Indeed, after breakup occurs, we see a recoil of the upper filament against gravity in Figure 2(c) due to surface tension and possible remnant elastic stresses inside the liquid.

At elevated temperatures above the gel point (75 °C), a solid with viscoelastic loss ratio $\tan\delta = G'' / G' \approx 0.4$ is implied from Figure 2(d). For this bounce test, yolk that has been held in a water bath at 75 °C for 1 hour is rolled into a ~1 cm diameter ball weighing ~0.5 g and dropped. With an initial drop height of $h_{\text{drop}} = 10$ cm and rebound height of $h_{\text{reb}} = 3$ cm, the loss tangent is $\tan\delta = \ln(h_{drop}/h_{reb})/\pi \approx 0.38$. This is associated with a specific timescale (frequency) and forcing stress. With a video camera shooting at 48 frames per second, impact occurs in less than $\Delta t = 1/48$ seconds, or at an impact frequency of at least $\omega = \pi/\Delta t \approx 150$ rad/s. After impact, the magnitude of the velocity changes by $\Delta v = 0.63$ m/s resulting in a characteristic shear stress of $\sigma = \rho R \Delta v/\Delta t \approx 150$ Pa The long time solid-like behavior is confirmed in Figure 2(e) with a 20 mL tilted vial; photo taken after 10 min.

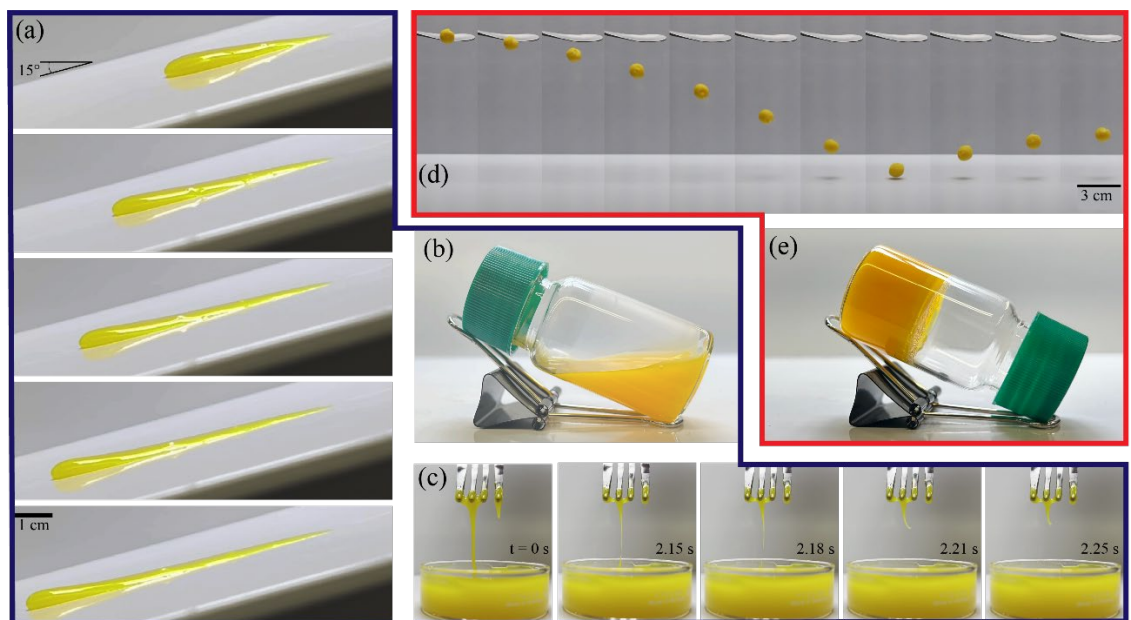

**Figure 2.** Protorheology tests reveal key rheological behavior below and above the gel temperature. (a)-(c) solution at 25 °C demonstrating (a) finite shear viscosity from flow down an incline (5 seconds between images), (b) liquid behavior taking the shape of the tilted vial (10 min after tilt), (c) extensional viscosity from capillarity-driven breakup of a fluid filament. (d)-(e) gel at 75 °C demonstrating (d) viscoelastic energy storage and loss from a bounce test (20 ms between images), (e) solid-like behavior from a high apparent viscosity with immobilized material at the top of a tilted vial (10 min after tilt).

At intermediate temperatures, yolks gradually transition from solution to gel, as shown in Figure 3 in the context of whole eggs. Here we placed a batch of room temperature whole eggs into a water bath initially heated to 55 °C and left them to equilibrate. After 1 hour, an egg was removed, cut, and photographed after which the temperature of the water bath was increased by 1 °C. The remaining eggs were left in the water bath for 25 min before another egg was removed and the process repeated. Preliminary tests showed that this 25-min equilibration time was long enough to ensure both the water bath and the egg yolk reached a spatially constant temperature and therefore microstructure. A panel containing photographs of each egg at temperatures between 55 °C and 74 °C is shown in Figure 3. Between room temperature and roughly 59 °C, the yolk behaves like a free-flowing liquid with little evidence of shear thinning or viscoelasticity observed. Between 62 °C and 63 °C the yolk begins to hold its shape when cut. Between 64 °C and 70 °C the yolk exhibits soft-solid-like yielding and at temperatures above 70 °C, the yolk exhibits a brittle-like fracture.

The images in Figure 3 suggest a gel point somewhere between 62 °C – 70 °C and provide context for the following detailed temperature-dependent study of linear and nonlinear viscoelastic properties. After identifying the critical gel condition in the following section, we then consider nonlinear rheometry and protorheology observations conducted near the critical gel point.

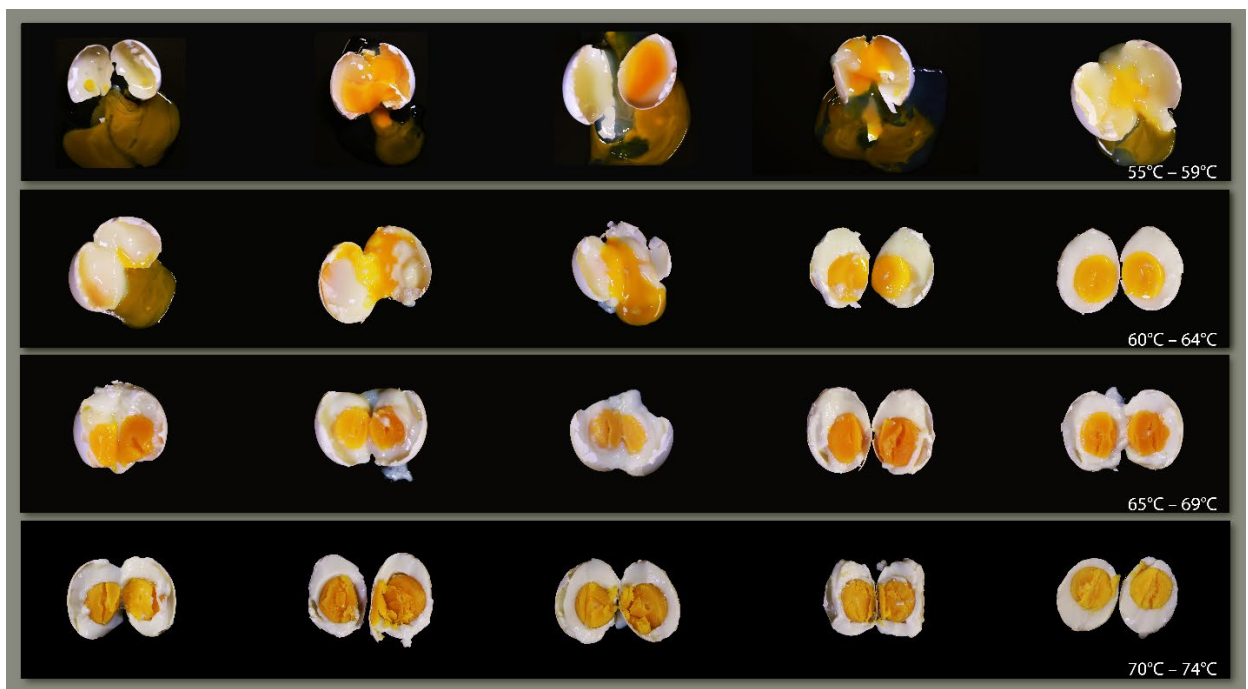

**Figure 3**. Gelation in the context of whole eggs: cross-section cuts after water-bath immersion at each temperature from 55 °C (top left) to 74 °C (bottom right) reveal the gradual transition through the gel point. All 20 eggs were placed in a water bath set to 55 °C and left for 1 hour. One egg was removed, the temperature increased by 1 °C, and the remaining eggs allowed to equilibrate for 25 min before the process repeated for every other egg  Pictures were taken roughly 10 seconds after the eggs were cut.

### B. Temperature-dependent linear viscoelasticity

Egg yolk is an excellent model system for realistic critical gels, but at some elevated temperatures, held at long times, yolk exhibits a continual slow time-dependent evolution (mutation) of microstructure and rheological properties. For example, Cordobés et al. (2004) reported that at 65 °C (using parallel plates on a rotational rheometer), the evolving viscoelastic moduli do not reach a precise plateau in their timescale of observation[12]. This realistic situation requires nuance and caution when determining a gelation temperature. To remedy this, some prior studies have *arrested* the gelation[12,51,52], which typically requires quenching the sample to a temperature below the gel temperature. A lower temperature alters the properties of the material and deviates from what we may experience in our own tactile and sensory interactions with cooked egg yolk. Therefore, we prefer to avoid this and conduct experiments at the temperature of interest, acknowledging that, though the material evolves in time, data can be collected quickly and in the same manner ensuring accurate comparisons across tests[53–55].

We observe the evolution of the egg yolk gel at two different temperatures. At 59 °C (Figure 4a), the evolution continues over an extremely long time – on the order of days. At 65 °C (Figure 4b), the evolution occurs on the order of just a few hours. We also coplot the digitized $G'$ data from Cordobés et al. (2004) for comparison, which shows generally good agreement despite using different geometries.

To determine whether there is a significant change in material properties, we use the dimensionless mutation number first defined by Winter et al.[55],

$$N_{mu} = \Delta t \frac{1}{G} \frac{dG}{dt} = \frac{\Delta t}{\tau_m} \quad (1)$$

where $\Delta t$ is the sampling time to collect data at a particular temperature, $G$ is a material property, in this case, modulus, and $\tau_m = G dt / dG$ is the mutation time: a characteristic time for the property to change by a significant amount. In our actual tests, we wait at least 10 min after a small change in temperature. As a worst case estimate for mutation rate, we use the elastic modulus data in Figure 4(b), which involves a large temperature jump. At 10 min, the mutation time for $G'$ is $\tau_m = 700$ s, which is a lower-bound estimate of the realistic $\tau_m$ in our other rheometric tests with smaller temperature steps. A single frequency sweep at fixed temperature requires $\Delta t = 410$ s, for which the mutation number is at most $N_{mu} = 0.58$. In fact, since our frequency sweeps were performed from 100 rad/s to 0.1 rad/s, the mutation number remains less than 0.5 (a reasonable limit) for the duration of the sweep except for the last point; i.e. the data at 0.1 rad/s takes 60 seconds to collect $(N_{mu} = 0.09)$. Our stress relaxation, creep compliance, and flow sweep tests last for 300 s, resulting in a slightly lower overall $N_{mu} = 0.43$. Since these are upper-bound estimates, we judge our realistic mutation numbers as small, $N_{mu} \ll 1$, which supports that we have a reasonably fixed state of the material at each temperature [9,55–58]. Tests at longer times would be at more risk of mutation artifacts. We therefore limit our 'long time' observations of liquid versus solid behavior to a few hundred seconds, which is also consistent with typical protorheology and culinary interactions with cooked egg yolk on the order of several minutes.

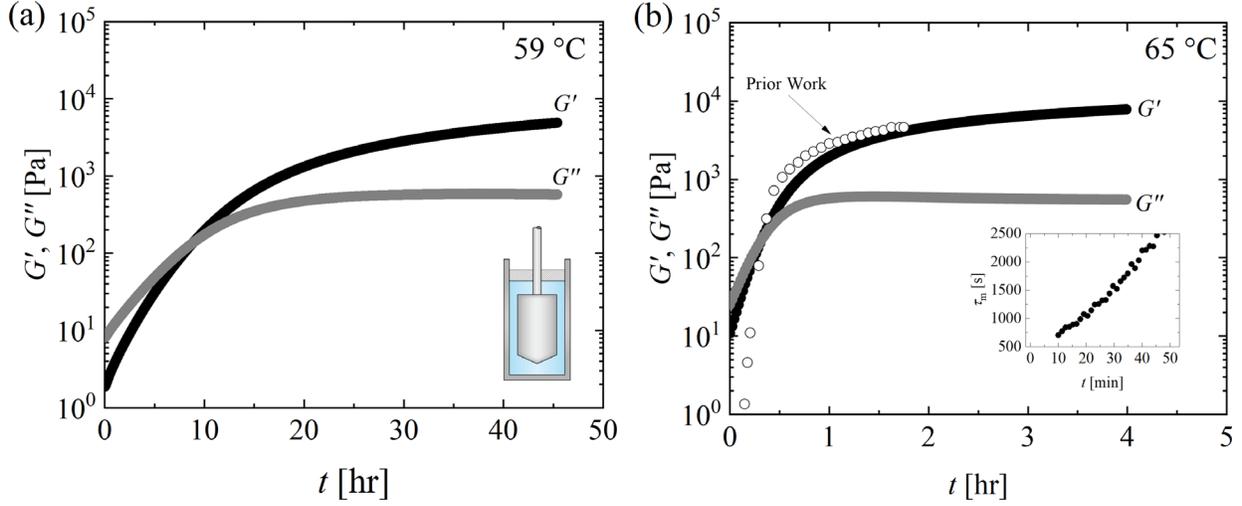

**Figure 4.** Time-dependent mutation of viscoelastic moduli after increase in temperature (fixed frequency 10 rad/s and strain amplitude 1%). (a) increase to 59 °C (b) increase to 65 °C. Both tests began at room temperature with a constant temperature ramp of 5 °C/min. Time $t = 0$ corresponds to when the Peltier system reached the target temperature. Geometry used was a 30 mm cup and 28 mm bob (schematic). $G'$ data from a similar procedure[12] is presented in (b) for comparison. Inset shows the mutation time, $\tau_m$ generated from $G'$ data in (b) after an initial temperature soak period: in this case, 10 min.

Despite the apparent simplicity of the critical gel point schematic in Figure 1, determining when the zero-shear viscosity diverges to infinity and the equilibrium modulus emerges from zero requires an infinite observation time[59]. In the all-too-common scenario where infinite observation time is impossible, one must use linear viscoelastic measures at finite times to determine the gel point. One common feature to use is power-law viscoelastic time dependence in the terminal relaxation regime. For example, in stress relaxation tests, the *gel equation*[8] describes the power law decay in shear modulus as

$$G(t) = St^{-n} \tag{2}$$

for which the equilibrium modulus

$$G_e = \lim_{t \to \infty} G(t) = 0 \tag{3}$$

while simultaneously the linear viscoelastic zero-shear viscosity diverges

$$\eta_0 = \int_0^\infty G(t)dt \to \infty . \tag{4}$$

This power law relaxation is therefore consistent with the critical point of solidification, for which $G_e = 0$ and $\eta_0 \to \infty$, as in the schematic of Figure 1. In oscillatory tests, Eq.(2) results in viscoelastic moduli having power-law frequency-dependence in the terminal regime, $G' \propto G'' \propto \omega^n$, and therefore a frequency-independent loss ratio, $\tan \delta = G''/G' = \text{const}$. This power-law viscoelastic signature is one way to identify the critical gel temperature, in addition to observing measures of zero-shear viscosity diverging with temperature, or elastic modulus emerging with temperature. We will consider each of these methods here.

It remains a common misconception that $G' = G''$ indicates the gel point, e.g. when moduli evolve with temperature as observed at fixed deformation frequency $\omega$, as in Figure 5 at two different frequencies. Note how the intersection point depends on frequency, which is true for most materials[60]. Furthermore, the moduli crossover temperature does not conform with our definition of the critical gel temperature as power-law viscoelasticity, or with a diverging zero-shear viscosity, or emerging equilibrium elastic modulus. Nevertheless, we can consider the crossover temperature as a type of "critical" temperature where a particular feature exists in the data. We refer to this temperature as the crossover temperature.

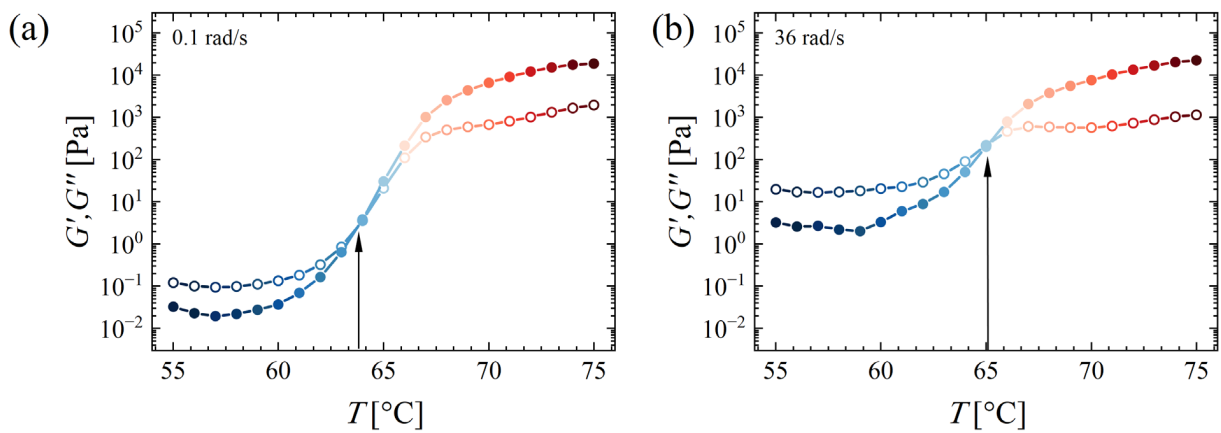

**Figure 5.** Crossover temperature depends on frequency of oscillation. Linear viscoelastic storage modulus (filled symbols) and loss modulus (open symbols) at a strain amplitude of 1%. (a) 0.1 rad/s, gives a crossover temperature of 63.6 °C (multimedia available online); (b) 36 rad/s gives a crossover temperature of 65.1 °C. (Data originally collected as frequency sweeps at each temperature.)

At both frequencies, at low temperatures the viscous response dominates, and we see a slight decrease in both moduli before increasing with temperature. This is consistent with prior work where a decrease in $G''$ is known to occur due to thermal agitation[12]. The decrease is more pronounced in the low frequency data which is due to thermal agitation affecting the viscous behavior of the yolk. For $\omega = 0.1$ rad/s (Figure 5a multimedia available online), a crossover is observed at 63.6 °C while at $\omega = 36$ rad/s (Figure 5b), a crossover is observed at 65.1 °C by interpolation. Although this is not a significant temperature difference for our egg yolk system, others have reported stronger frequency-dependence of the crossover point[61,62]. The moduli as a function of temperature for each frequency tested are presented in Appendices A 1 – A 3 to observe the shifting crossover temperature.

The moduli crossover method is practical to identify a critical temperature, but it is inconsistent with the diverging zero-shear viscosity and emerging equilibrium modulus that define the gel point, as in Figure 1. From the same frequency sweep dataset, terminal regime power-law behavior can be used as a surrogate for infinite time of observation[59,63], as can emergence of an apparent equilibrium modulus or diverging zero-shear viscosity.

Figure 6 shows the frequency-dependent viscoelastic moduli for different temperatures. In Figure 6a (multimedia available online), we present the complex moduli at three representative temperatures: below, near, and above the gel point. We also plot the instrument inertia limits as a dashed line[38]. At 55 °C, we do not observe $G' \sim \omega^2$ and $G'' \sim \omega$ indicating we are not in the terminal regime. At 64 °C, we begin to see a plateau in the low frequency $G'$ signal (an indicator of solid behavior) while higher frequencies do not show power law behavior of parallel $G'$ and $G''$. It is critical to note that this low frequency plateau can also be caused by mutation of the sample. Since frequency sweeps were done from high frequency to low frequency, data points collected at low frequencies would not only be collected after the sample has mutated, but the low frequency points take a longer time to collect, further obscuring this effect. In Figure 6b and Figure 6c, we observe some regions of power law behavior for temperatures below and above 64 °C making the power-law signature potentially imprecise when determining the gel point for this available range of frequency. Isolated frequency sweeps at each temperature are provided in Appendices A 4 – A 5 to observe the shifting power law behavior.

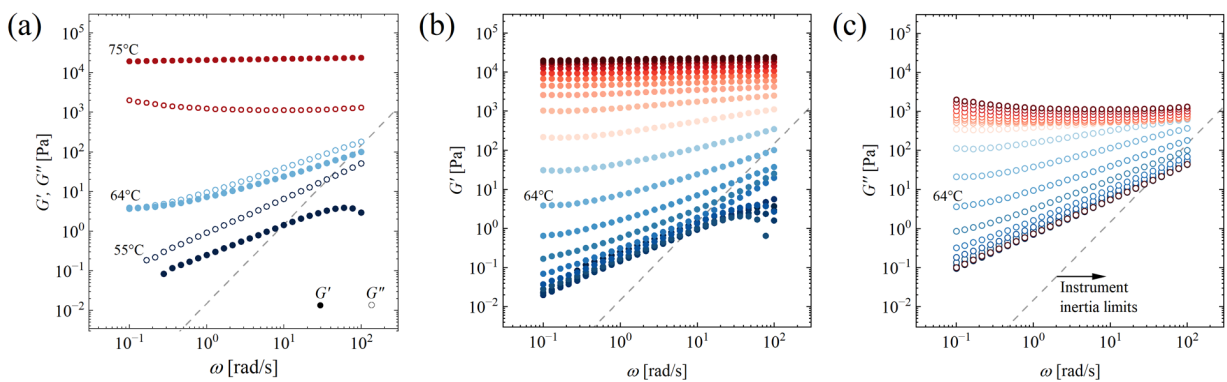

**Figure 6.** Temperature-dependent frequency sweeps from 55 °C – 75 °C. (a) three representative temperatures: below, near, and above the critical gel temperature. At 64 °C, a low-frequency $G'$ plateau appears (multimedia available online); (b) storage moduli for all temperatures; (c) loss moduli for all temperatures. Power-law frequency-dependence is evident at temperatures above and below 64 °C. An instrument inertial limit is shown as a dashed line in each plot[38].

Power-law critical gel behavior is identified in Figure 7, as the temperature where frequency-independent tan $\delta(\omega,T)$ occurs. Here we plot the loss tangent tan $\delta(\omega,T)$ as a function of temperature for our entire range of frequencies, $\omega = 0.1 - 100$ rad/s. In this projection, the frequency-independent tan $\delta$ visually appears as what we refer to as the "pinch-point". In our data, this signature of power-law viscoelasticity occurs at $T_{c,\text{PL}} = 66.3$ °C (interpolated). Also coplotted is the digitized tan $\delta$ data from Cordobés et al. (2004), the best prior dataset for egg yolk gelation. Those authors measured $G'$ and $G''$ with a fixed heating rate of 0.75 °C/min at frequencies of 0.63, 3.1, 6.3, and 14 rad/s[12]. Using a different protocol with different geometry, they reported a region of frequency-independent tan $\delta$ behavior from 66.3 °C – 66.8 °C. The crossover temperatures, $T_x$, from Figure 5 are equivalent to tan $\delta = 1$ in Figure 7, which occurs over a range of temperatures depending on the frequency. The "pinch-point" for the power-law critical gel occurs at a temperature above that range. At this power law critical gel temperature, we observe $G''$ to be roughly 40% of $G'$, in contrast to the $G'' = G'$ at the crossover temperatures. At low temperatures, increasing frequencies correspond to increasing values of tan$\delta$ and at high temperatures, increasing frequencies correspond to decreasing values of tan$\delta$. The frequency dependence is explicitly depicted in Figure 9.

While the $\tan\delta$ "pinch-point" is more rigorous than a moduli crossover point, it does not explicitly consider the asymptotic behavior of long-observation-time properties highlighted in Figure 1. The power-law behavior is associated with a particular range of observation timescales only up to about 10 seconds associated with the lowest frequency. In fact, we can see signatures of gelation in the terminal regime such as the plateau in $G'$ in Figure 6 at low frequency, beginning to appear at temperatures below $T_{c,PL}$. This motivated our consideration of longer-time observation from stress relaxation and creep compliance tests.

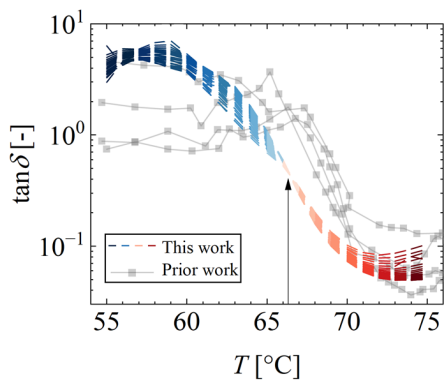

**Figure 7.** Critical gel temperature from $\tan\delta(T;\omega)$ frequency-independence (ranging from 0.1 – 100 rad/s, i.e., timescales 0.01 – 10 s). Projecting onto the two-dimensional $\tan\delta(T)$ axes reveals a visual "pinch point" at a critical temperature of 66.3 °C. For context, a prior study is shown which reported $\tan\delta$ at 0.63 – 14 rad/s and determines a range of frequency independence at temperatures between 66.3 °C – 66.8 °C[12].

Stress-relaxation and creep-compliance tests up to 300 seconds are shown in Figure 8. In a linear stress-relaxation test, power-law slope in modulus is indicative of critical gel behavior. In the terminal regime, this is the only type of decay which will result in an infinite viscosity but finite modulus. In a linear creep-compliance test, steady viscous flow is indicated by a power-law slope of 1 on log-log axes. As the material gels, the slope will become progressively smaller while remaining power-law.

In Figure 8a, we plot linear stress-relaxation data where the arrow indicates the direction of heating. Power law decay in the modulus is first observed at 63 °C, with much less noise at 64 °C. At 65 °C, we fit Eq.(2) to the $G(t)$ curve between 50 milliseconds and 100 seconds. The fit parameters are gel strength $S = 51$ Pa.s$^{0.42}$ and power law index $n = 0.42$. Above the gel temperature, we would expect the modulus to eventually plateau at long-times to the equilibrium modulus. Instead, we see an initial power law decay followed by a much more rapid decay in the modulus after ~100 s, as if power-law viscoelasticity is developing but it is not representative of a terminal regime. Recognizing our finite time of observation, to obtain an effective apparent equilibrium modulus, we averaged the value of the modulus over the last 30 seconds for each test between 66 °C – 75 °C.

In Figure 8b, we plot linear creep compliance data where the arrow indicates the direction of heating. A deviation from a power-law slope of 1 in the terminal regime first occurs at 63 °C. To determine the effective apparent zero-shear viscosity of the yolk, at our longest timescales of observation, we averaged the viscosity values over the last 30 seconds, until 64 °C and above. At higher temperatures, the compliance began to reach a plateau. At 67 °C and above, we averaged the last 30 seconds of the compliance data and took the inverse to report the effective apparent equilibrium shear modulus.

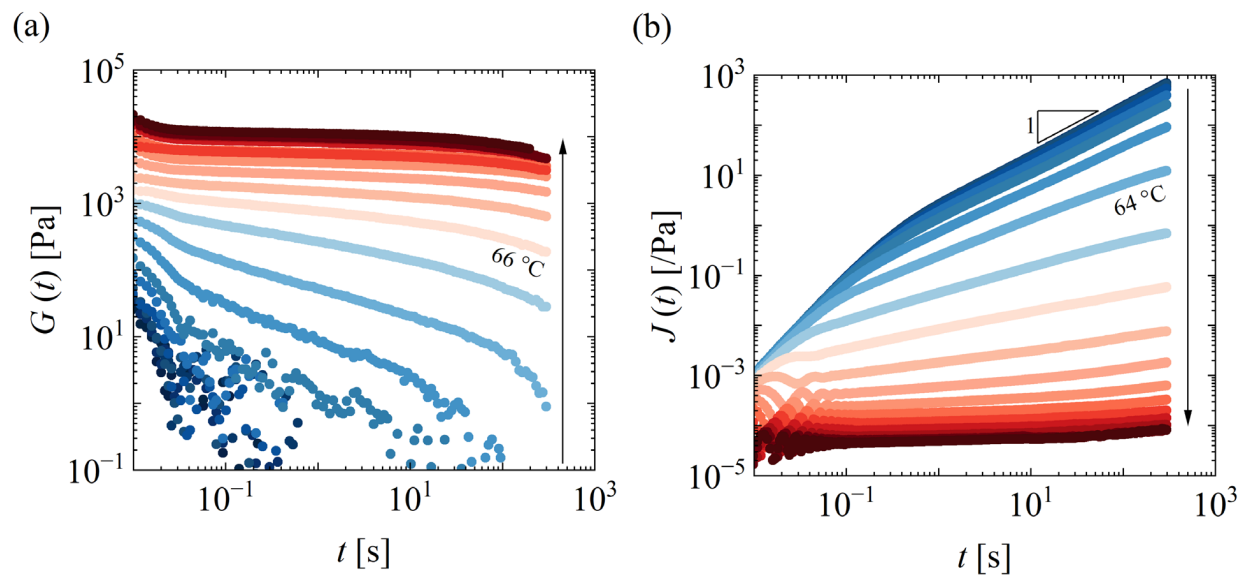

**Figure 8.** Step tests to observe longer times up to 300 s. Temperature range 55 °C – 75 °C (arrow in the direction of heating). (a) Stress relaxation modulus from step strain input ($\gamma_0$ = 1%); equilibrium modulus extracted at 66 °C and above. b) Creep compliance strain evolution from step stress input ($\sigma_0$ = 0.1 Pa and 2 Pa); steady shear viscosity extracted up to and including 64 °C.

Figure 9 summarizes the four ways in which we determine critical temperatures of egg yolk gelation. The most common technique, approximation by moduli crossover (Figure 9a), is a quick experiment to run but the crossover temperature depends on the frequency of oscillation. To instead determine a gel temperature based on diverging zero-shear viscosity and emerging elastic modulus, from the same dataset in Figure 9a, we plot the dynamic viscosity and storage modulus at the lowest observed frequency, as shown in Figure 9b. Around the gel temperature, we use the equations[57]

$$\eta_0(T) \sim (T_c - T)^{-s}; \qquad \text{for } T_c < T \qquad (5)$$

$$G_e(T) \sim (T - T_c)^z; \qquad \text{for } T < T_c, \qquad (6)$$

to fit to the diverging viscosity and emerging modulus. Fitting Eq.(5) to the dynamic viscosity between 62 °C and 65 °C results in the asymptotic gel temperature $T_{c,\eta'}$ = 68.8 °C. Likewise, fitting Eq.(6) to the storage modulus between 65 °C and 69 °C results in the asymptotic gel temperature $T_{c,G'}$ = 64.9 °C. We decided that fitting to 5 datapoints was sufficient to accurately capture the behavior and these particular temperature ranges provided the best $R^2$ value. The gel temperatures we obtain are associated with a particular timescale of observation $t = 1/\omega = 10$ seconds, for which the critical temperature $T_c$ is between 64.9 °C and 68.8 °C.

For direct comparison, in Figure 9c we again plot the loss tangent as a function of temperature. The "pinch-point" defines a critical temperature $T_{c,PL} = 66.5$ °C. This is larger than any of the crossover temperatures but within the range of $T_{c,\eta'}$ and $T_{c,G'}$. Finally, in Figure 9d, we present the viscosity and modulus measurements obtained from stress relaxation (Figure 8(a)), creep compliance (Figure 8(b)), and steady shear (Figure 10), each of these defined by long time observations in response to step inputs (step strain, step stress, and step strain rate, respectively). From data in Figure 8a, we can directly plot an equilibrium shear modulus $G_e$, and from Figure 8b we can plot both zero-shear viscosity $\eta_0$ and equilibrium shear modulus $G_e$. The steady strain rate data for $\eta_0$ comes from a following section but is plotted here as well. We observe generally good agreement in the viscosity values between both methods. However, the modulus data has a significant discrepancy between methods, especially near the gel temperature. Using Eq.(5) and Eq.(6) again, we fit to 5 data points to produce 4 different critical temperatures between 64.3 – 65.8 °C. We consider each of these asymptotes as separate methods to identify a critical temperature and a solidification point, each associated with a particular timescale of observation and type of forcing applied to the material.

The gel temperature from each method and its associated timescale are detailed in Table 1. Crossover temperatures are included for comparison, but we do not consider these defensible definitions of a gel point. The other critical temperatures may be considered as gel points, as they are determined by either directly observing material properties at long times or by observing power-law behavior as a surrogate for long-time of observation. In general, the longer that one is able to observe, the lower (earlier) the gel temperature. Longer observation times means a larger viscosity can be measured while still remaining a liquid, thus a steeper diverging zero-shear viscosity. Similarly, the longer the time of observation, the lower the measurable modulus can be while still remaining finite.

Among the many possible definitions of gel point, the diverging zero-shear viscosity from the creep tests $T_{c,\eta(\sigma)}$ may have the most significant impact on observable flow physics, e.g., with our tactile interaction and protorheology observations; stress conditions are easier to enforce than strain conditions. In the following section we therefore focus on T = 65 °C as a representative temperature near the critical gel point where zero-shear viscosity has diverged.

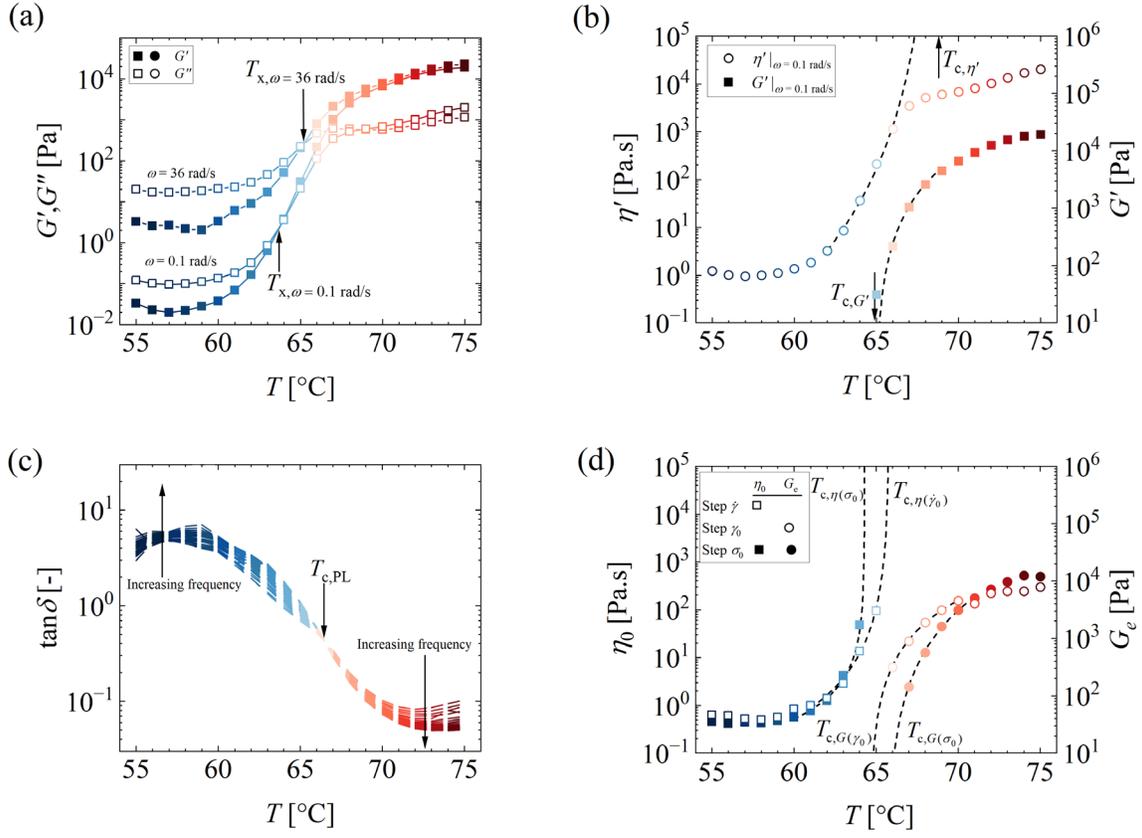

**Figure 9.** Comparison of critical temperatures from different methods. (a)-(c) oscillatory tests down to 0.1 rad/s, i.e., up to $t_{obs}$ = 10 s, (d) step inputs up to $t_{obs}$ = 300 s. Specific values of critical temperature summarized in Table 1. (a) moduli crossover temperatures, $T_x$, increase with imposed frequency; (b) replotting data as dynamic viscosity $\eta' = G''/\omega$ and storage modulus $G'$ reveals more fundamental asymptotic behavior and a range of critical temperatures. Dashed lines are fit to 5 data points using Eq.(5) and Eq.(6). (c) $\tan\delta(T;\omega)$ "pinch-point" for power law viscoelasticity temperature as in Figure 7. (d) step input to observe longer time behavior and a range of critical temperature ($T_{c,\eta(\sigma)} < T_c < T_{c,G(\sigma)}$).

Table 1. Summary of critical temperatures and their associated observation timescales from the methods detailed in Figure 9.

| Name | Variable | Value [°C] | Timescale | Comments |
|---|---|---|---|---|
| Crossover temperature | $T_{x,\omega=0.1\,rad/s}$ | 63.6 | $t = 10$ s | Moduli crossover is insufficient to determine solid versus liquid behavior, but can serve as an approximate metric |
| | $T_{x,\omega=36\,rad/s}$ | 65.1 | $t = 0.03$ s | |
| Gel temperature | $T_{c,PL}$ | 66.3 | $t = 0.01 - 10$ s ($\omega = 0.1 - 100$ rad/s) | Frequency-independent tan $\delta$ is evidence of viscoelastic power-law time dependence but using a range of frequencies is a surrogate for infinite time of observation |
| | $T_{c,\eta'}$, $T_{c,G'}$ | 64.9 – 68.8 | $t = 10$ s ($\omega = 0.1$ rad/s) | Asymptotic divergence and emergence of dynamic properties at a fixed frequency |
| | $T_{c,\eta}$, $T_{c,G}$ | 64.3 – 65.8 | $t_{long} = 300$ s | Asymptotic divergence and emergence at longer observation times after step input; longest observation may be limited by mutation of the sample or patience of the observer |

## C. Nonlinear rheology and protorheology near the critical gel temperature

Much of the flow we impose during cooking and eating is nonlinear such as large amplitude shear or extension. To probe this behavior, we begin with nonlinear steady shear across the entire range of temperatures. The ramp down viscosity for each temperature is shown in Figure 10 with the shaded region indicating the low torque limit of the instrument[38]. The inset is the viscosity versus stress which indicates dramatic yielding for temperatures above 67 °C. We fit a Carreau model,

$$\eta(\dot{\gamma}) = \eta_\infty + (\eta_0 - \eta_\infty)\frac{1}{\left[1 + (K\dot{\gamma})^2\right]^{(1-n)/2}} \tag{7}$$

to each curve up to and including 65 °C, after which, the fit no longer converged. The zero-shear viscosity values from the Carreau fits are plotted in Figure 9d. As the temperature increases, so does the viscosity and its slope approaches -1 indicating a development to yield stress fluid. Above 71 °C however, the viscosity decreases at low stress, consistent with being an effective solid in the linear viscoelastic regime.

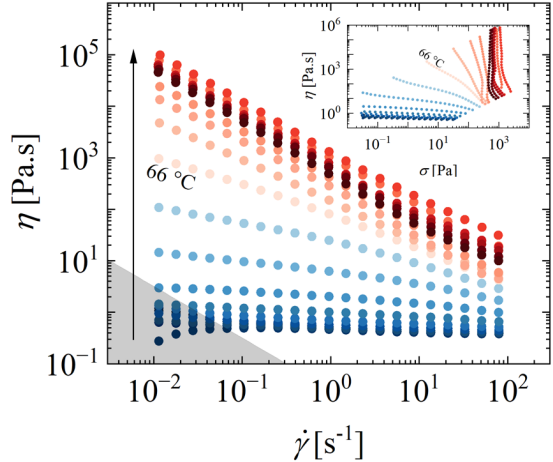

**Figure 10**. Steady shear flow for temperatures 55 °C – 75 °C (arrow in the direction of heating). Shear rate ramps down from 100 s$^{-1}$ to 0.01 s$^{-1}$ at each temperature reveal a transition from weakly non-Newtonian to yield-stress fluid across the gel point. The inset shows the same data as viscosity versus stress indicating dramatic yielding behavior for temperatures above 67 °C. A Carreau model is fit to the data at each temperature to determine the zero-shear viscosity. Above 65 °C, the fit no longer converges. The gray region shows the limit of low torque with a safety factor of 20.[38]

Of particular interest is large amplitude oscillatory shear (LAOS) near the gel point shown in Figure 11. We report the first harmonics of storage and loss moduli in Figure 11(a). The response shows a weak strain overshoot consistent with many soft glassy materials[64] and a strain softening consistent with yield-stress fluids, e.g., prior work on Laponite colloidal gels near the critical point[65]. Sweeping back down in strain amplitude shows that the response is almost completely reversible with only a ~50% smaller $G'_1$ value at intermediate strain amplitudes during sweep down. A slight increase in $G'$ after sweeping down is due to the sample mutating. To diagnose possible slippage at the wall or shear banding, we considered testing at different gaps. However, we are avoiding parallel plates for reasoning mentioned above (free surface artifacts) and we cannot change the gap with our concentric cylinder geometry. As an alternative, we consider that wall slip or shear banding can reveal itself as nonmonotonic stress or a sharp plateau of stress. We plot the stress amplitude as a function of strain amplitude in Figure 11b. The monotonic stress growth supports the assumption that slip or shear banding are not occurring.[66]

Immediately after completing the experiment described above, we collected data in transient mode to ensure the sample did not change much between tests and collect full waveform data. Figure 11c shows the elastic-projection Lissajous curves obtained from sweeping up to 1%, 10%, and 100% strain. Qualitatively we see a transition from elastic behavior at low strain to soft gel behavior at larger strain. Many metrics have been developed to quantify the LAOS response[64,67–71]. In this study we report the rotation, distortion, and plastic dissipation ratio ($R$, $D$, and $\phi$) as relevant low-dimensional metrics to describe the complex LAOS response. These are defined as:

$$R = \left| \frac{G'_1 - G'}{G'} \right| \tag{8}$$

$$D = \left|\frac{G'_L - G'_M}{G'_M}\right| \quad (9)$$

$$\phi = \frac{\pi \gamma_0 G''_1}{4\sigma_{max}} \quad (10)$$

where $G'_1$ and $G''_1$ are the first harmonics of the storage and loss moduli, $G'_L$ is the large-strain modulus or tangent modulus and $G'_M$ is the minimum-strain modulus or secant modulus. Calculating the parameters as above, we obtain $R = 0.88$, $D = 0.99$, and $\phi = 0.63$. The plastic dissipation ratio helps to quantify that significant underlying elasticity still exists even at this large strain amplitude of 100%.

Locating this material on a rotation-distortion map, we can compare to other protein-based gels, as in Piñeiro-Lago et al. (2023)[69], their Figure 6. We see that egg yolk critical gel is unique compared to the available data in that figure. Compared to Afuega'l Pitu cheese (20 °C), critical gel egg yolk has much more distortion of Lissajous curves at the same rotation. Compared to the protein-polysaccharide mucin gel (slug slime), egg yolk critical gel has much more rotation at the same distortion. This indicates an interesting amount of intracycle material nonlinearity. Whether other critical gels have similarly unique nonlinear LAOS rheology remains an interesting open question to explore. What we can say conclusively is that compared to other foods, such as the cheeses reviewed in Piñeiro-Lago et al. (2023)[69], their Figure 9, critical gel egg yolk is softer than the softest cheese. This might be expected and reinforces our intuition of these properties: the elastic modulus around 600 Pa, and critical stress around 60 Pa, are off the charts lower than available LAOS data for even the softest cheeses.

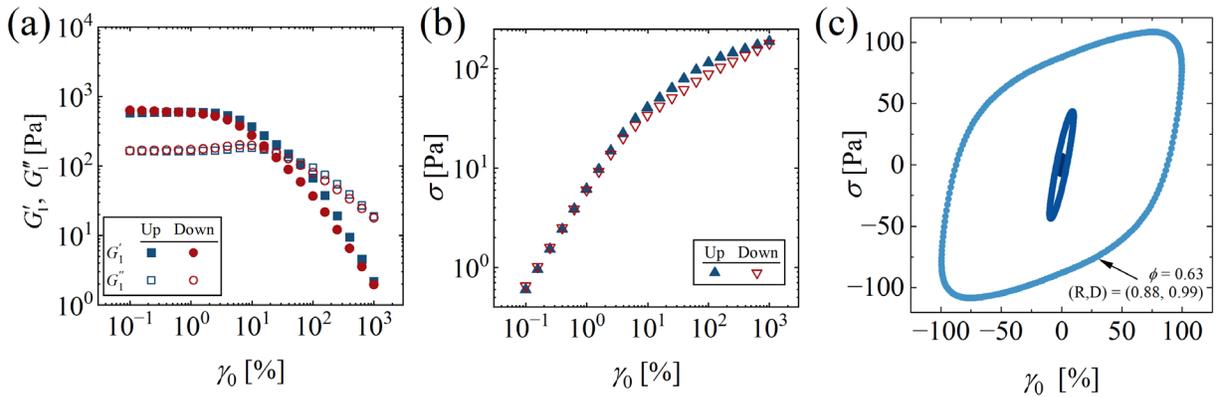

**Figure 11**. LAOS tests show significant nonlinearity above 100% strain. (a) amplitude sweeps up and down between 0.1% strain and 1000% strain show good reversibility as (b) the pure stress versus strain amplitude shows a monotonic growth ruling out wall slip. (c) Elastic Lissajous-Bowditch curves at 1%, 10%, and 100% strain amplitude show a significant distortion and rotation nonlinearity.

We also probed the extensional behavior of the yolk near the gel point and compared it to egg yolk at ambient and at elevated temperatures. Figure 12 shows pictures of egg yolk between two 8 mm diameter plates initially 4 mm apart and suddenly separated at a constant nominal true strain rate (Hencky strain rate) of $\dot{\varepsilon} = 0.5 \text{ s}^{-1}$. The extensional strain to break[72] is similar between the 25 °C yolk and the 65 °C yolk at around 60% strain. At 75 °C, the yolk exhibits adhesive failure rather than cohesive failure. At long times $(t = 60 \text{ s})$, the yolk at room temperature returns to a viscous pile rounded off by surface tension. For the yolk around the gel point, the material on the bottom plate maintains most of its features indicating yield-stress like behavior. However, immediately before breaking, a thin filament forms in the room temperature yolk indicating slight extensibility of the egg yolk proteins. Additionally, the 65 °C yolk at long times begins to slump indicating either a time-dependent or a low yield stress. This behavior near the critical gel point is similar to some extensible yield stress fluids[72,73].

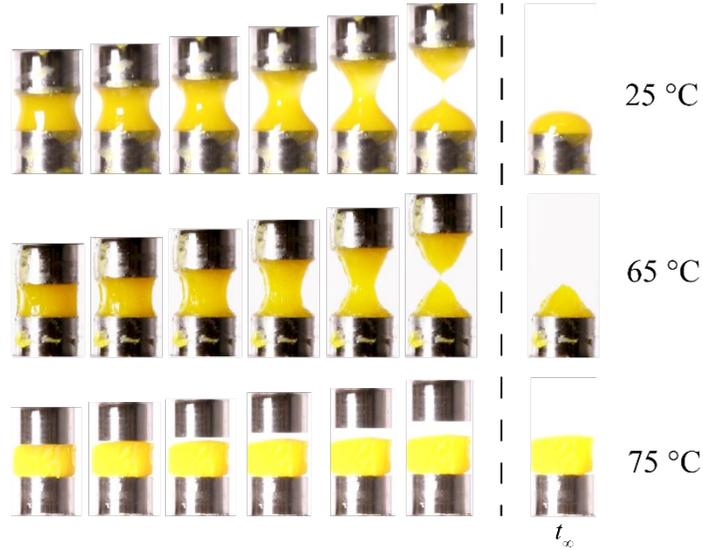

**Figure 12.** Imposing a constant nominal true strain rate of $\dot{\varepsilon} = 0.5 \text{ s}^{-1}$ below, near, and above the gel temperature reveals three different modes of failure. At 25 °C, a thin liquid filament bridge is formed before breakage. At 65 °C a shorter thin filament is formed, and the failure mode is more brittle. At 75 °C, the material exhibits adhesive failure from the top plate. The final picture (at $t_\infty$) taken 60 seconds after failure.

We can directly experience the linear and nonlinear rheology of critical gel egg yolk (protorheology) and use this to understand aspects of measurable rheometric properties. Here we select specific flow and deformation scenarios that are accessible to the reader, visually observable, and related to measurable properties.

Figure 13 shows pictures of tests we call "jiggle", "spread", and "splat" along with their accompanying rheological signatures revealed in the flow physics. In Figure 13(a) (multimedia available online), a ~1 cm thick sample of yolk around the critical gel temperature is adhered to the side of a vertical plate. The plate is then struck with a hammer causing inertio-elastic oscillations of the yolk mass. More specifically, a viscoelastic wave propagates through the yolk. With a video camera recording at 120 frames per second, immediately after impact, the yolk begins to oscillate with a period of ~4 frames or $T = 0.03 \text{ s}$. Assuming that thickness of the sample is a half-wavelength $\lambda \approx 2 \text{ cm}$, the wave propagates with a wave speed $c = \lambda / T \approx 0.7 \text{ m/s}$. The frequency of oscillation is $\omega = 2\pi/T = 200 \text{ rad/s}$. With a density of ~1 g/cm$^3$, an estimate of the shear modulus is $|G^*| = \rho c^2 \approx 500 \text{ Pa}$ [28] at these conditions. The yolk deforms by roughly ~1 mm at the free surface producing an elastic recoverable strain of $\gamma_0 \approx 10\%$.

The characteristic driving stress is therefore $\sigma = |G^*|\gamma_0 \approx 50$ Pa. After providing an initial impulse, the wave dampens out in 14 frames or 3.5 cycles. Being elastic enough to oscillate, but with high damping, is a key aspect of critical gel rheology, e.g., the high damping at high frequency in Figure 13(d).

In Figure 13b (multimedia available online), we place yolk on a flat surface. No flow is observed after 10 seconds. After pressing down with a spatula, it takes 2 seconds to spread the yolk 1.7 cm. With a yolk thickness of ~3 mm, the resulting apparent strain rate is $\dot{\gamma}_a \approx 3$ s$^{-1}$. We estimate that an upper bound on the spreading force is 1 N (the weight of a medium-sized apple). This acts over a spatula area of $\sim 6 \times 10^{-4}$ m$^2$ resulting in a shear stress of $\sigma \approx 1700$ Pa and a resulting upper bound on viscosity of $\eta \approx 600$ Pa.s. This demonstrates a ductile and dramatically shear thinning behavior, as that of a yield-stress fluids. Indeed, the rheological properties revealed by this accessible protorheology test are consistent with the steady shear flow data in Figure 13(e).

In Figure 13c (multimedia available online), we drop a ~1 cm radius scoop of yolk ($\rho \approx 1$ g/cm$^3$) weighing ~4g and collect pictures of the trajectory every 5 ms. With an initial height of 500 cm, the velocity upon impact is ~3 m/s. The "splat" occurs over 4 frames (15 ms) after which it remains adhered to the surface. With no bounce ($\Delta v \approx 3$ m/s), the characteristic stress is $\sigma = \rho R \Delta v / \Delta t \approx 400$ Pa. In Figure 13(f), this regime is dominated by the $G''$ response resulting in large dissipation. During impact, we see a flattening of the yolk followed by an elastic recoil to the final state. Despite the large dissipation, the yolk has a high recoverable elastic strain at these applied stresses. This is likely distinct to the critical gel, since at higher temperatures the yolk becomes more brittle (Figure 1).

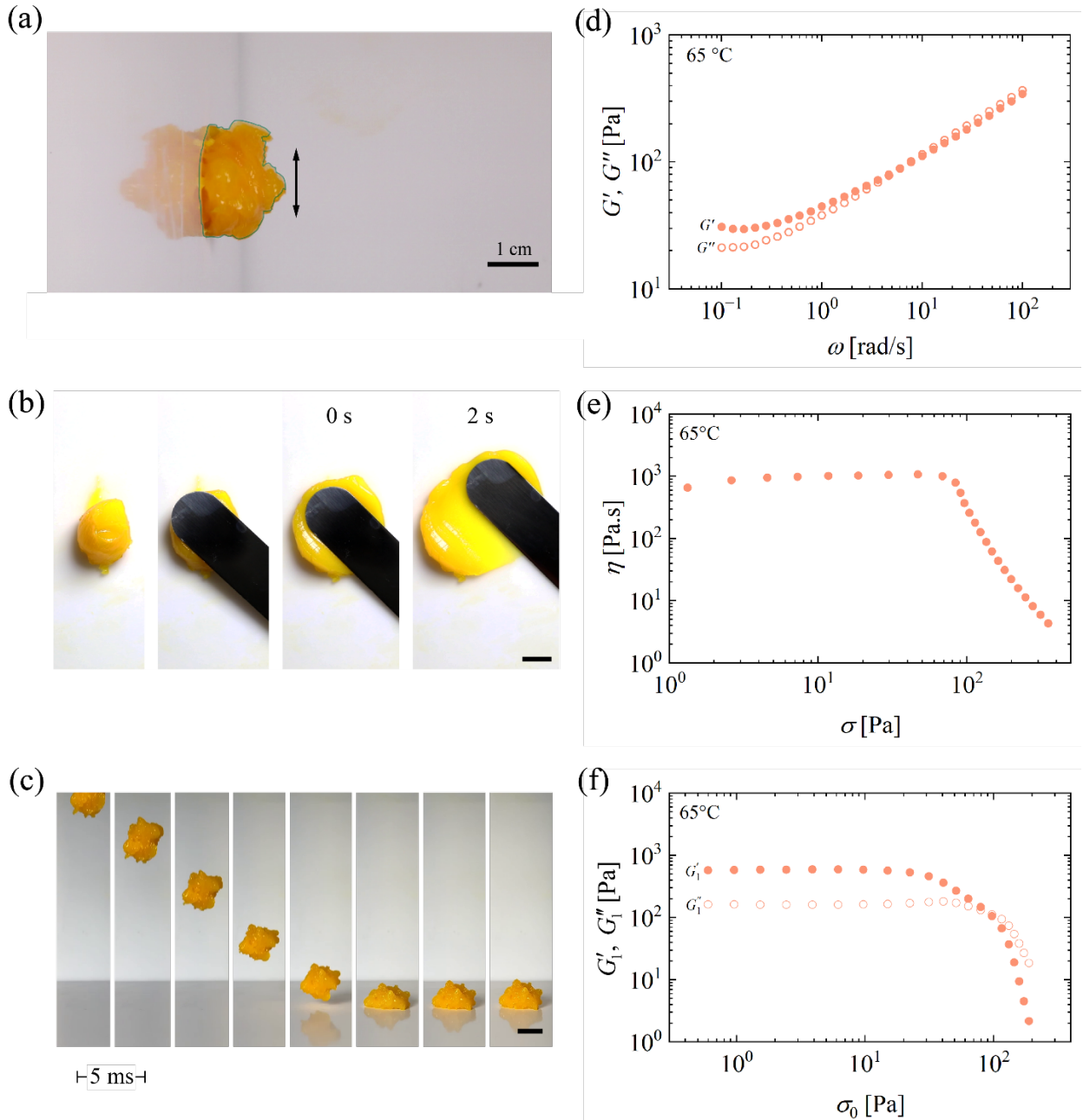

**Figure 13**. Protorheology tests near the gel point reveal (a) some damping after small amplitude short time impact (multimedia available online), (b) spreadability under large shear deformation (multimedia available online), and (c) significant damping after large impact shear combined with compression (multimedia available online). The relevant rheometer data near the critical gel point is include in (d) – (f). All scale bars 1 cm.

## V. Conclusions

In this paper, we explored the rheology of a critical gel providing context both before and after gel transition. We chose an egg yolk for its prevalence around the globe and its accessibility to those with limited lab experience. This non-toxic edible material provides accessible insight into the physics of a critical gel, in particular the impact of critical gel rheology on observable flow behavior. In many ways, an egg yolk is a model critical gel. However, it also reveals many critical challenges in rheometry. First, above a certain temperature, egg yolk is not thermally stable and will mutate over time making transient tests difficult to trust. Second, the yolk was observed to form a film on parallel plate geometries – even with the addition of mineral oil – necessitating the use of concentric cylinders with a thin film of oil floated on top. Third, the LAOS tests revealed that the gel network is not fully reversible above a critical forcing amplitude making repeat measurements difficult.

Our results present a full complement of both rheometric measurements (dataset available as Supplementary Information) and corresponding high-quality images (also available as Supplementary Information) of egg yolk protorheology. By integrating protorheology photos and videos with rigorous rheometric data, we provide a deeper understanding of the physics of the critically important behavior of critical gels, with broader impacts for teaching and modeling sol-gel transitions.

## Supplementary Material

We provide the complete dataset in the supplementary material as a downloadable excel file.

## Acknowledgements

The authors thank Sarah Wu for art assistance and acknowledge support for this work as part of the Regenerative Energy Efficient Manufacturing of Thermoset Polymeric Materials (REMAT), an Energy Frontier Research Center funded by the U.S. Department of Energy, Office of Science, Basic Energy Sciences at the University of Illinois Urbana-Champaign under Award No. DE-SC0023457.

Appendix

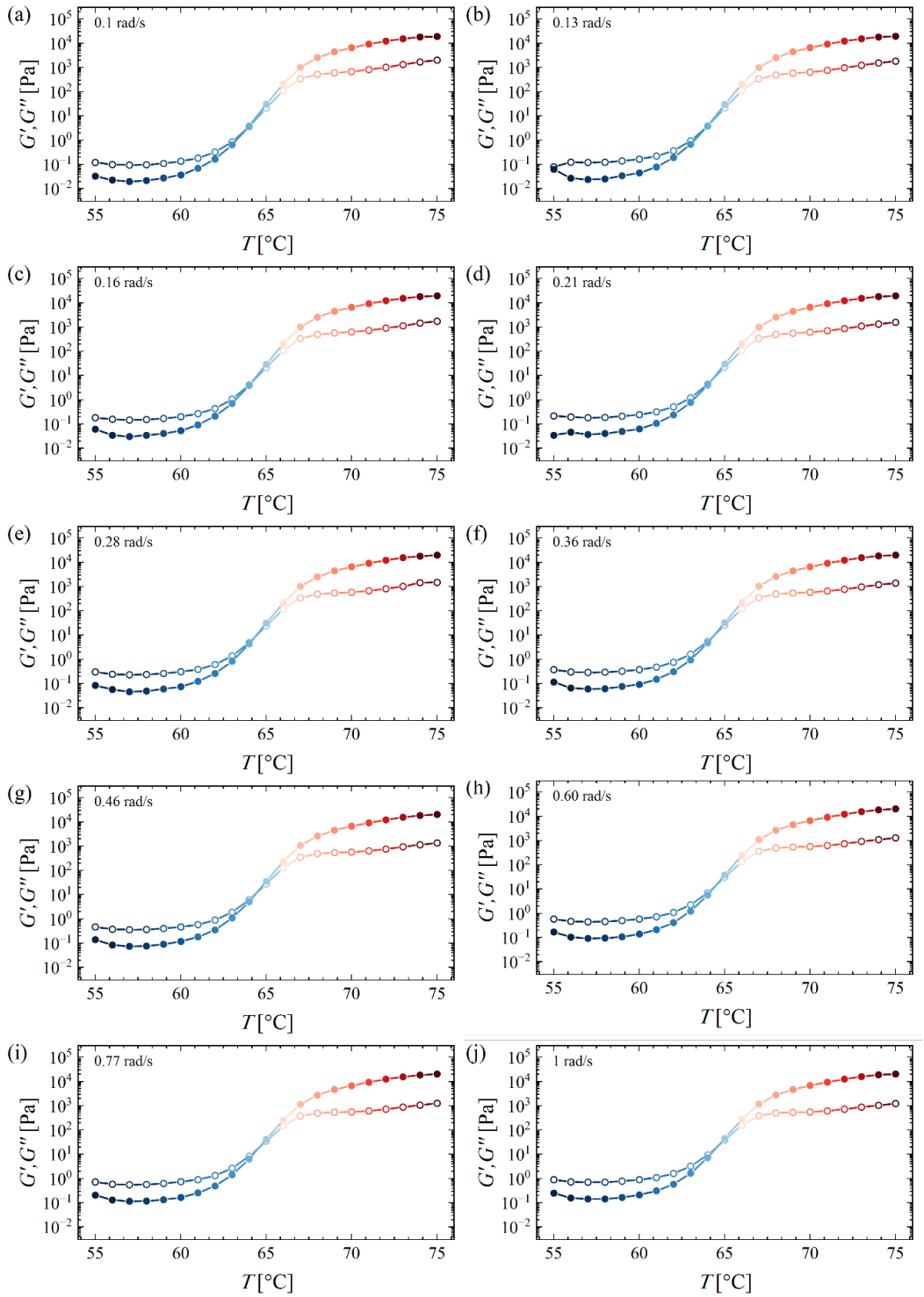

A 1. Moduli as a function of temperature for each frequency tested between 0.1 rad/s and 1 rad/s.

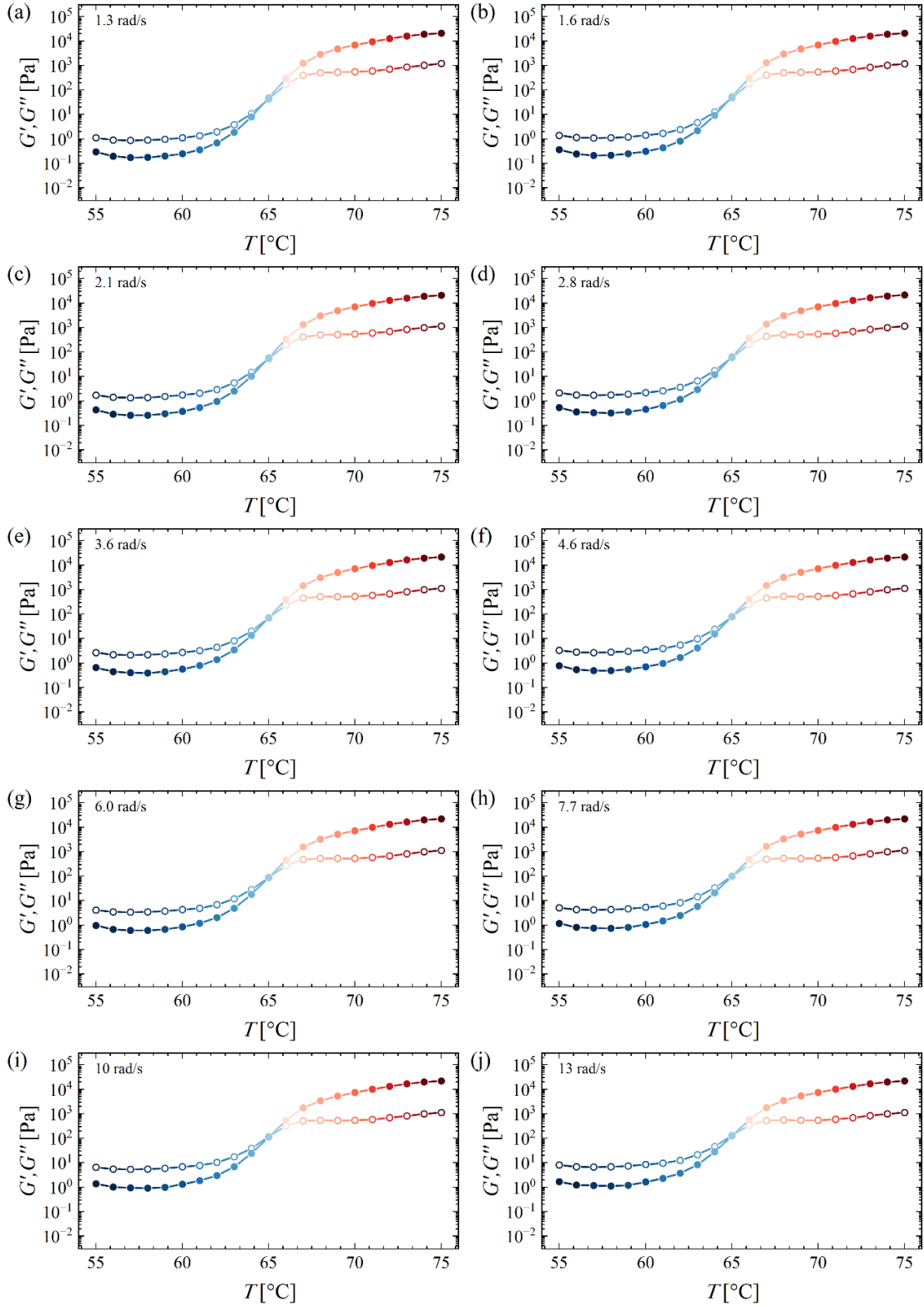

A 2. Moduli as a function of temperature for each frequency tested between 1.3 rad/s and 13 rad/s.

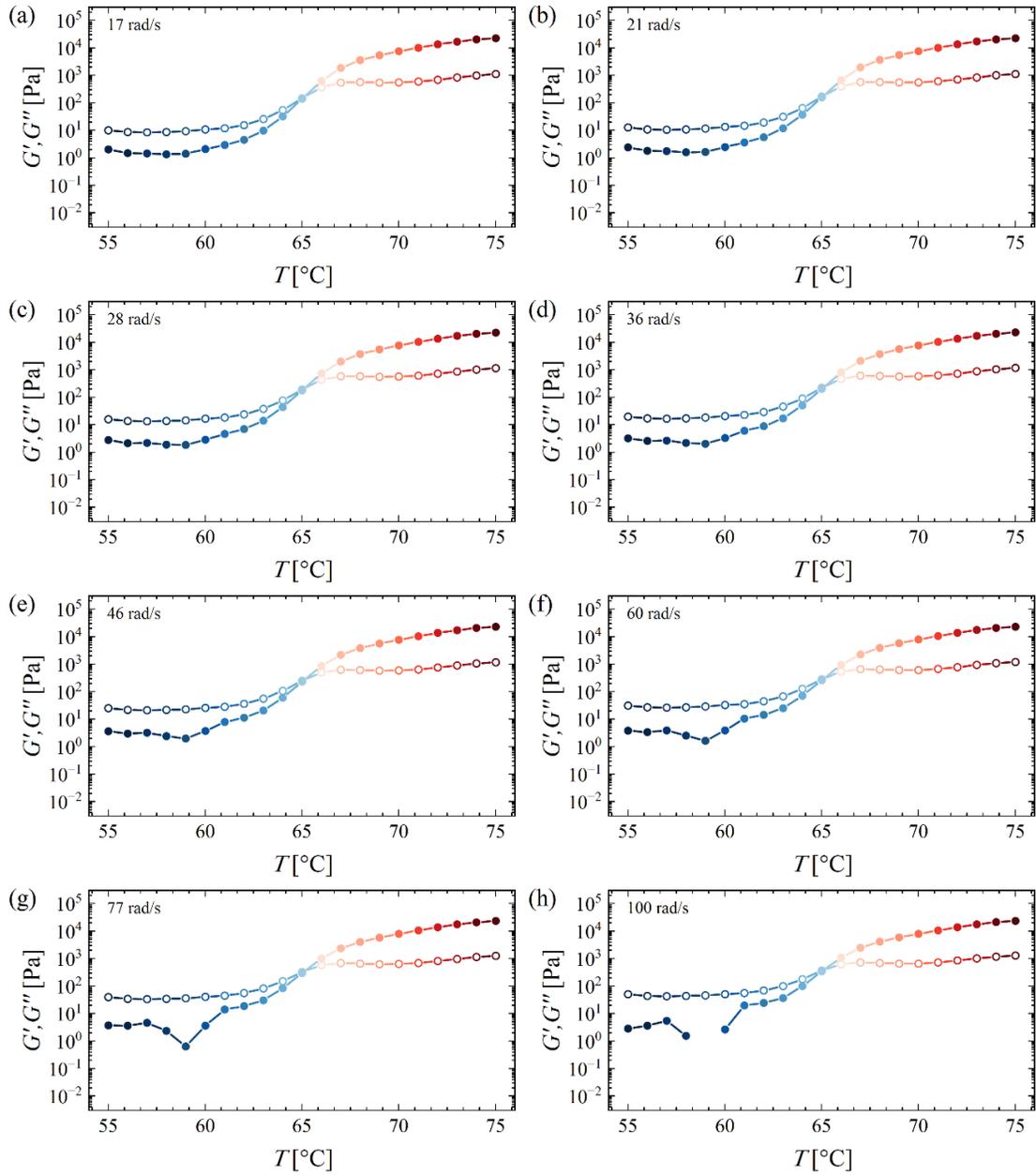

A 3. Moduli as a function of temperature for each frequency tested between 17 rad/s and 100 rad/s.

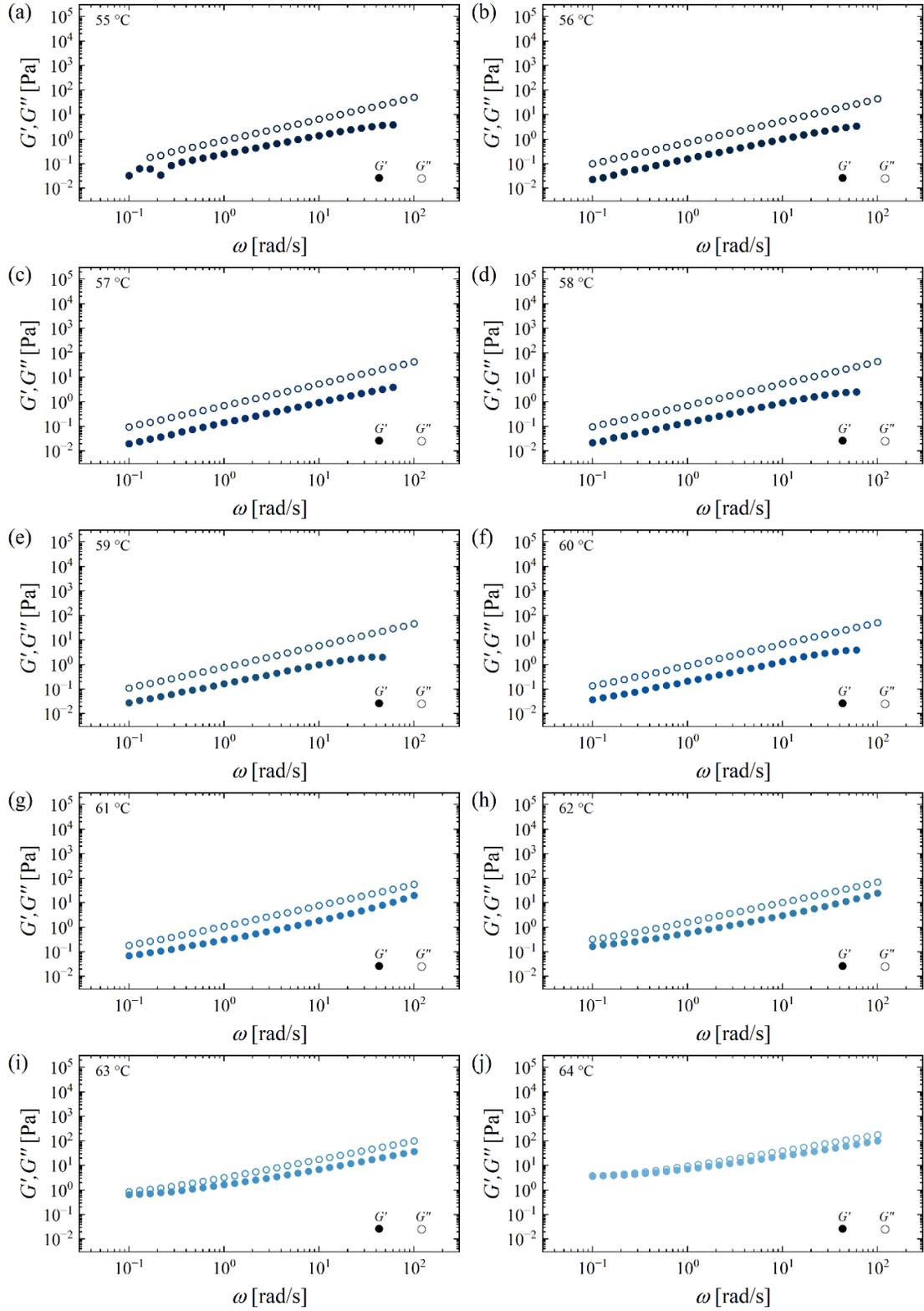

A 4. Moduli as a function of frequency for each temperature tested between 55 °C and 64 °C.

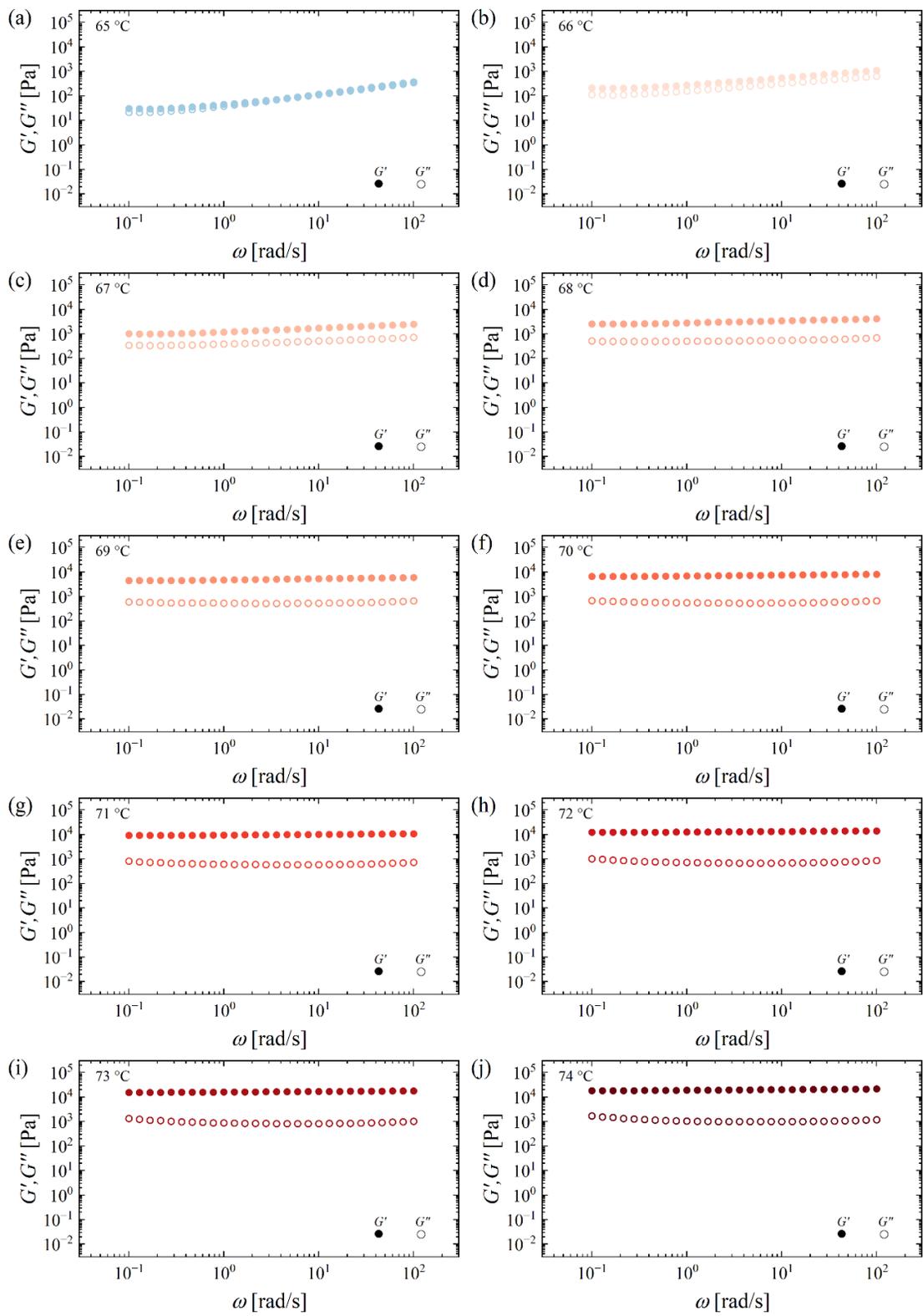

A 5. Moduli as a function of frequency for each temperature tested between 65 °C and 74 °C.